\documentclass[prd,showpacs,superscriptaddress,twocolumn,nofootinbib]{revtex4}

\usepackage{amsfonts}
\usepackage{amsmath}
\usepackage{amssymb}
\usepackage{amsthm}
\usepackage{bm}
\usepackage{dcolumn}
\usepackage{epsfig}
\usepackage{graphicx}
\usepackage{graphics}
\usepackage[latin1]{inputenc}
\usepackage{latexsym}
\usepackage{rotating}
\usepackage[dvips]{color}

\newcommand\be{\begin{equation}}
\newcommand\ba{\begin{eqnarray}}
\newcommand\ee{\end{equation}}
\newcommand\ea{\end{eqnarray}}

\newcommand{\BD}{{\mbox{\tiny BD}}}

\newcommand{\GW}{{\mbox{\tiny GW}}}
\newcommand{\GR}{{\mbox{\tiny GR}}}

\begin{document}
\title{Constraining the evolutionary history of Newton's constant \\ with gravitational wave observations}

\author{Nicol\'as Yunes}
\affiliation{Department of Physics, Princeton University, Princeton, NJ 08544, USA.}

\author{Frans Pretorius}
\affiliation{Department of Physics, Princeton University, Princeton, NJ 08544, USA.}

\author{David Spergel}
\affiliation{Department of Astrophysical Sciences, Princeton University, Princeton, NJ 08544, USA.}
\affiliation{Princeton Center for Theoretical Science, Princeton University, Princeton, NJ 08544, USA.}

\date{\today}

\begin{abstract}
Space-borne gravitational wave detectors, such as the proposed {\emph{Laser Interferometer Space 
Antenna}}, are expected to observe black hole coalescences to high redshift 
and with large signal-to-noise ratios, rendering their gravitational waves ideal probes of fundamental 
physics. The promotion of Newton's constant to a time-function introduces modifications to 
the binary's binding energy and the gravitational wave luminosity, leading to corrections in the chirping 
frequency. Such corrections propagate into the response function and, given a gravitational wave observation,  
they allow for constraints on the first time-derivative of Newton's constant at the time of merger. 
We find that space-borne detectors could indeed place interesting constraints on this quantity as a function
of sky position and redshift, providing a {\emph{constraint map}} over the entire range of redshifts where binary black 
hole mergers are expected to occur.  A LISA observation of an inspiral event with redshifted 
masses of $10^{4}-10^{5}$ solar masses for three years should be able to measure $\dot{G}/G$ at the time of merger 
to better than $10^{-11} \; {\rm{yr}}^{-1}$.
\end{abstract}


 


\maketitle

\section{Introduction}
The principle of equivalence is at the cornerstone of classical physics as evidenced by the opening paragraph
of Newton's {\emph{Principia}}~\cite{1760pnpm.book.....N}. Constructing a relativistic theory of gravity
consistent with this principle was one of the key inspirations that led Einstein to general 
relativity (GR)~\cite{1916AnP...354..769E,2005AnP....14S.303E,Will:1993ns,lrr-2006-3}. 
The Strong Equivalence Principle (SEP) states that {\emph{all}} physics must satisfy:
(i) the Weak Equivalence Principle (WEP), 
(ii) local Lorentz invariance and (iii) local position invariance.
The WEP states that test-particle trajectories are independent of their internal structure,
while the latter two properties require experimental results to be independent of the 
$4$-velocity and position of the laboratory frame.

Many known alternative theories of gravity violate the SEP~\cite{lrr-2006-3}\footnote{The one known exception is Nordstr\"om's 
conformally-flat scalar theory~\cite{Nordstrom.comf.flat}, which has been ruled out by 
light-deflection Solar System observations~\cite{lrr-2006-3}.}, 
a classic example of which are scalar-tensor theories~\cite{Will:1993ns}. These theories modify the Einstein-Hilbert Lagrangian by 
multiplying the Ricci scalar by a field $\phi$ and adding a kinetic and a potential term (in the Jordan frame). 
Scalar-tensor theories are well-motivated from a theoretical standpoint, as they seem unavoidable in unification models
inspired by particle-physics. In string theory, for example, there are numerous (dilatonic) scalar fields that couple
to the scalar curvature, and perhaps also to the matter sector, then leading to additional violations of the 
WEP~\cite{1988PhLB..213..450T,Maeda:1987ku,Damour:1994zq,Damour:2002mi,Damour:2002nv}. 
In loop quantum gravity, the Barbero-Immirzi parameter that controls the minimum eigenvalue of the area operator
could be promoted to a field, leading to a classical coupling of Einstein gravity with a scalar-field stress-energy 
tensor~\cite{Taveras:2008yf,Mercuri:2009zt}.  

From an observational standpoint, we also possess evidence for interesting physics that could point 
to a modification of GR on cosmological scales, the prime example of which is the late-time acceleration of the universe.
The cause of this acceleration is usually interpreted within the framework of pure GR as ``dark energy'', a
new form of matter that is or behaves similar to a cosmological constant. This form of matter
has only recently (redshifts less than of order unity) become important in governing the evolution of 
the universe. Another interpretation of the observed acceleration is that it is purely a gravitational
phenomenon, though the theory governing the dynamics of spacetime is not GR, but a new
or modified theory that has cosmological solutions that mimic a cosmological constant within pure GR.
It is therefore of some interest to search for modifications to GR, in particular as a function of
redshift.

Often, gravitational theories that violate the SEP predict a varying gravitational constant.
In the scalar-tensor theory case, one can effectively think of the scalar as promoting 
the coupling between gravity and matter to a field-dependent quantity via $G \to G(\phi)$, thus violating local position invariance 
as $\phi$ varies. Another interesting example are higher-dimensional, brane-world scenarios, 
where although physics remains Lorentz invariant in higher dimensions, ``gravitational leakage'' into the bulk  
inexorably leads to a time-varying effective 4D gravitational constant~\cite{Deffayet:2007kf}. A constraint on the
magnitude of this variation can also be interpreted as a constraint on the curvature radius of 
extra dimensions~\cite{Johannsen:2008tm}.

Existing experiments searching for varying $G$
can be classified into two groups: (a) those that search for the present or nearly present rate of variation (at redshifts close
to zero); (b) those that search for {\emph{secular}} variations over long time periods (at very large redshifts). The most stringent bounds
of the first class come from lunar ranging observations~\cite{Williams:2004qba} and pulsar timing observations~\cite{Kaspi:1994hp,Deller:2008jx},
but also include planetary radar-ranging~\cite{2005AstL...31..340P} and surface temperature observations of low-redshift  
millisecond pulsars~\cite{Jofre:2006ug,Reisenegger:2009cq}.
The strongest bounds of the second class come from Big Bang Nucleosynthesis (BBN) calculations~\cite{Copi:2003xd,Bambi:2005fi}, 
but also include the evolution of the Sun~\cite{1998ApJ...498..871G}. The order of magnitude of the strongest constraint from either class is about $\dot{G}/G \lesssim 10^{-13} \; {\rm yr}^{-1}$, but it varies with the experiment.  

In this article, we show that gravitational wave (GW) observations of binary compact object
coalescence can also provide constraints on the variation of $G$. 
We concentrate on these coalescences
because of their high expected event rate and signal-to-noise ratio (SNR) with the proposed
{\emph{Laser Interferometer Space Antenna}} (LISA)~\cite{lisa}. 
We consider both nearly equal-mass, supermassive black hole (SMBH) merger events 
(with total mass $M\sim 10^{6} M_{\odot}$), as well as extreme-mass ratio, quasi-circular 
inspirals of a stellar-mass black hole (BH) into a SMBH. 
Only BHs are considered here because they lack internal structure that could lead to additional violations
of the WEP~\cite{lrr-2006-3} and modifications of the multipolar structure of the gravitating body.
From this standpoint, BH binaries are {\emph{cleaner}} probes of fundamental physics. 

One interesting difference between previous tests of $\dot{G}$ and GW-based ones is that we are using
black holes, ie.~{\emph{vacuum}} solutions, instead of neutron stars. As mentioned earlier, scalar-tensor theories can be interpreted 
as variable-G theories, but in that case the variable $G$ is really only a variation in the coupling between 
gravity and matter. 
The gravitational constant $G$, however, serves more fundamentally as the quantity that
defines the relationship between geometry (length) and energy, and such a quantity is
{\em not} altered in scalar-tensor theories. We here consider the possibility 
that this fundamental constant varies with time. In this sense, we are looking at more generic variable-G theories, instead of restricting
attention to theories that can reduce to a scalar-tensor theory. 
In Appendix~\ref{appendix}, we discuss the relation between generic variable-G theories 
and scalar tensor ones in more detail.

Another interesting difference between previous tests of $\dot{G}$ and the GW tests proposed 
here is that the latter constrain $\dot{G}/G$ at {\em{the time and location of the merger event}}. 
Therefore, once a sufficient number of events have been observed, a {\emph{constraint map}} 
could be constructed, which would encode bounds on $\dot{G}/G$ along our past light-cone, ie.~as a function of redshift $z$ 
and sky position, out to a redshift of order $10$ and limited mostly by the likelihood of merger events at that redshift. 
In contrast, the first class of tests discussed above can only bound $\dot{G}/G$ near redshift zero, while the BBN calculations 
only constrain linear changes in $G$ from $z=0$ to $z \gg 10^{3}$ and they are degenerate with limits on the number of 
relativistic species. GW constraint maps would thus be excellent probes 
of local position invariance and the SEP, complementing previous bounds in a regime where the latter are inaccessible. 

Constraint maps are {\emph{generic}} and applicable to {\em any} parametrizable GR deviation in the waveform, 
provided one possesses GW observations from sources at sufficiently large distances. 
The concept of GW tests, of course, is not new (see eg.~\cite{Will:1994fb,Will:1997bb,Scharre:2001hn,Finn:2001qi,sutton:2002:bgm,
Dreyer:2003bv,Berti:2004bd,Berti:2005qd, Will:2004xi,Alexander:2007kv,Yunes:2008bu,Arun:2009pq,Stavridis:2009mb,Yagi:2009zm} 
for a discussion of GW tests of Brans-Dicke and massive graviton theories), but until now constraint maps had not been really considered. 
Such maps add a new dimension to the nature of GW tests; this is in contrast to most traditional tests that can only search
for deviations in our immediate spacetime neighborhood, or search for particular deviations that are by fiat constant over the age of the Universe.
In this sense, GW constraint maps can provide invaluable global information on the SEP and local position invariance that is unattainable 
with Solar System observations. 
 
We find that a variability in Newton's constant introduces modifications to the phase of the GW response function of an 
inspiraling binary that could be measurable by LISA to an interesting level. 
For comparable-mass BH inspirals with total redshifted mass $M_{z} = 10^{6} M_{\odot}$, one could constrain  
${\dot{G}}/G \lesssim 10^{-9}  \; {\rm{yr}}^{-1}$ or better to redshift $10$ (using an SNR of $10^3$~\cite{Baker:2006kr}),
while for extreme mass-ratio inspirals with the same total mass, 
one could place roughly the same constraint out to redshift $0.3$ (using 
an SNR of $10^2$~\cite{AmaroSeoane:2007aw}). 
Such a bound becomes stronger for intermediate mass BH inspirals, allowing constraints on the order of $10^{-11} \; {\rm{yr}}^{-1}$.
Even though these bounds are, at a first glance, weaker than those from existing tests, the latter only constrain secular variation in $G$, 
or oscillatory changes if we happen to be near a maximum of ${\dot{G}}$ today. GW observations are in principle sensitive
to a much larger class of variations in $G$, if a sufficient number of events are detected to build a meaningful constraint map.

We also find that there is an equivalence between the vacuum, varying-G model that we consider here and a time-varying 
chirp mass. Such an equivalence is established via the mapping $(\dot{G}/G) {\cal{M}} \to \dot{\cal{M}}$. In view of this, 
one can think of the tests presented here as probes of the time-constancy of the BH mass or horizons. Such time-variation
could be induced, for example, due to gravitational leakage into the bulk in braneworld models~\cite{Johannsen:2008tm}. 
Therefore, the results presented here are truly a fundamental GW test of the gravitational interaction. 

An outline of the remainder of the paper is as follows. 
Section~\ref{sec_var} describes the parameterization chosen for the functional dependence Newton's constant.
Section~\ref{sec_mod} derives how the inspiral is modified in post-Newtonian (PN) theory with a time-varying $G$ 
to ${\cal{O}}(1/c^{3})$, where $c$ is the speed of light (this is commonly referred to as an expansion to $1.5$PN order).
Section~\ref{sec_cons} estimates the accuracy to which $\dot{G}/G$ could be measured by LISA, performing a
simplified Fisher analysis and a time-domain dephasing study, using both frequency-domain PN waveforms
and modified EOB, numerical templates. 
Section~\ref{sec_conclusion} concludes with some discussion of future extensions of this work.
In Appendix~\ref{appendix}, we further discuss the mapping between scalar-tensor theories and a variable-G theory, 
while in Appendix~\ref{appendix-cosmo} we comment on possible degeneracies due to cosmological expansion.

Henceforth, we use the following conventions: we work exclusively in
four spacetime dimensions with signature $(-,+,+,+)$~\cite{Misner:1973cw}; 
Greek letters $(\mu,\nu,\ldots)$ in index lists range over all spacetime indices; 
Latin letters $(a,b,\ldots)$ in index lists range over all template parameter indices $\lambda^{a}$; 
overhead dots stand for partial differentiation with respect to the time coordinate,
$\dot{E} := \partial E/\partial t$; the Einstein summation convention is employed unless 
otherwise specified; finally, we use geometrized units $G=c=1$ where possible, with $c$ the speed
of light in vacuum, but we shall mostly retain factors of Newton's constant $G$, as this quantity will be 
promoted to a time-function.

\section{Parameterization of the Variability of Newton's Constant}
\label{sec_var}

In constructing the GW signal, we do not want to restrict attention to a particular alternative theory. Rather,
we will {\em assume} that, during inspiral, the modification can be captured by simply promoting $G$ to a function 
of time in the {\em GR equations of motion}: $G \to G(t)$. This is analogous to promoting $G$ to a time-function
in Newton's second law, a topic much studied in the past~\cite{1937Natur.139..323D,1988NuPhB.302..645W,1989RvMP...61....1W,2003RvMP...75..403U}. 
The $G$-varying framework we consider here can thus be
thought of as an {\emph{effective theory}} that could represent {\emph{any}} alternative theory that leads to such a modification. 
In this section we discuss how we parameterize $G(t)$. 

At a given merger event, we will simply describe the evolution of the gravitational constant by
\be
\label{parametrization}
G(t,x,y,z) \approx G_{\rm c} + \dot{G}_{\rm c} \left(t_{c} - t\right), 
\ee
which is a Taylor expansion of $G(t,x,y,z)$ to first order in time $t$ and position $(x,y,z)$ about the coalescence event 
$(t_{c},x_{c},y_{c},z_{c})$, where $G_{\rm c} = G(t_{c},x_{c},y_{c},z_{c})$, and $\dot{G}_{\rm c} = (\partial G/\partial t)(t_{c},x_{c},y_{c},z_{c})$.
Such an expansion is only valid provided the spatial variation in $G$ is much smaller than the temporal variation, 
and the characteristic period of the temporal variation is longer than the observation window, the latter of which 
is on the order of a few years for LISA. 

Previous bounds on $G(t)$ have assumed linear
deviation from the {\em present day} value. In turn, this allowed BBN calculations to place such
stringent constraints on $\dot{G}$, by exploiting the huge ``lever arm'' provided by
integrating over cosmic time. However, any such bound will be largely insensitive to 
oscillatory components in $G$ that have periods much less that the cosmic observation time. 
With the parameterization chosen in Eq.~\eqref{parametrization}, such oscillations can be
captured, provided their frequency is much smaller than the inverse of the LISA observation time. 

Before proceeding, one might wonder how a variable $G$ can lead to modifications to a {\emph{vacuum}} spacetime, 
such as a BH inspiral. Newton's constant appears explicitly on the right-hand of Einstein's equations, 
$G_{\mu\nu} = 8\pi G/c^5 \ T_{\mu\nu}$, as a coupling constant between geometry, 
encoded by the Einstein tensor $G_{\mu\nu}$, and matter, given by the stress energy tensor $T_{\mu\nu}$.
If $T_{\mu \nu} = 0$, as is the case in vacuum, then one might naively conclude
that a variable $G$ would not introduce any modification. In fact, this is essentially the case in scalar-tensor theories
(see Appendix~\ref{appendix}), where the no-hair theorem is protected in vacuum by the scalar field's evolution equation.

Newton's constant, however, plays a much more fundamental role than merely a coupling constant: it defines the relationship 
between energy and length. For example, in the {\em vacuum} Schwarzschild solution, the relationship between the radius $R$ of the black hole 
and the rest-mass energy $E$ of the spacetime is $R=2 G E/ c^4$. Similarly, in a binary BH spacetime, the BHs 
have introduced an energy scale into the problem that is quantified by a specification of Newton's
constant\footnote{In a full numerical implementation, one could
effectively model a time-varying $G$ by introducing a time-varying length-scale, which is most easily done via a 
time-varying speed of light $c(t)$. One must point out, however, that such modifications are unrelated to
the variable speed of light theories of~\cite{Moffat:1992ud,Clayton:1998hv}.}. 
In our PN implementation of a varying $G$, these issues are avoided as $G$ explicitly couples to a point-particle
stress-energy tensor (the PN representation of a BH), and thus, it explicitly appears in the equations of motion.
Henceforth, we treat the variable-G modifications as induced by some effective theory and we do not concern
ourselves with the specific model from which such modifications arise. 

\section{Post-Newtonian Expansion of the Modified GW Signal}
\label{sec_mod}

In the above parameterization of $G(t,x,y,z)$ we have assumed that the spatio-temporal variation of $G$ is 
sufficiently long to only induce a secular evolution over the duration of the observed inspiral and merger.
Thus, the effect on the waveform will build up slowly over many cycles. We can then ignore the relatively short 
coalescence and ringdown phases of the merger to a good approximation, and concentrate only in the inspiral.
Modifications to the inspiral waveform can then be analyzed within PN theory; here
we shall carry all expansions out to ${\cal{O}}(1/c^{3})$ (ie.~$1.5$PN order). 

The time-dependence of Newton's constant modifies the generation of GWs due to corrections to the balance law during
inspiral. Consider then a binary BH system, represented by PN point-particles with binding energy  $E =-\eta M c^{2} x/2$, 
where $\eta = m_{1} m_{2}/M^{2}$ is the symmetric mass ratio, $M = m_{1} + m_{2}$ is the total mass, with
$m_{1,2}$ the binary's component masses, and $x = (2 \pi G M F/c^{3})^{2/3}$, with $F(t)$ the orbital 
frequency. The balance law states that the rate of change of the binary's binding energy $\dot{E}$ is equal and opposite to 
the GW luminosity ${\cal{L}}_{\GW} = 32 c^{5} \eta^{2} x^{5}/(5 G)$, where the latter can be obtained from Isaacsson's stress-energy 
tensor~\cite{Isaacson:1968ra,Isaacson:1968gw}. The promotion of $G \to G(t)$ in the conservative energy or Hamiltonian 
is comparable to a modification to Newton's second law, while this promotion in the GW flux is a modification
to the dissipative sector of the theory.  

With the parameterization of Sec.~\ref{sec_var}, the balance law can be rewritten
as the following, modified differential equation for the orbital frequency: 
\ba
\dot{F} &=& \frac{\dot{G}_{\rm c}}{G_{\rm c}} F + \frac{48}{5 \pi c^{5}} \left(G_{\rm c} {\cal{M}}\right)^{-2} x^{11/2} \; \eta^{11/5}
\nonumber  \\ 
&\times&
\left\{1 - \left( \frac{743}{336} + \frac{11}{4} \eta \right) x + 4 \pi x^{3/2}  
\right.  
\label{Fdot-Eq}
\\ \nonumber
&-& \left. \frac{5}{3} \frac{\dot{G}_{\rm c}}{G_{\rm c}} \left(t-t_{c}\right) \left[1 - \left(\frac{743}{240} + \frac{77}{20} \eta\right) x 
+ \frac{32}{5} \pi x^{3/2} \right]
\right\},
\ea
where ${\cal{M}} = \eta^{3/5} M$ is the chirp mass. By construction, this expansion is valid to $1.5$ PN order, 
ie.~for orbital velocities $v \ll c$ to relative ${\cal{O}}(v^{3})$. This expression can be simplified by using the zeroth-order
in $\dot{G}_{\rm c}$ relationship between time and frequency:
\ba
t_{\GR}(F) &=&  t_{c}  - \frac{5}{256} G_{\rm c} {\cal{M}} x_{\rm c}^{-4} \eta^{-8/5}
\left[ 1 
\right. 
\nonumber \\
&+& \left. 
\left(\frac{743}{252} + \frac{11}{3} \eta \right) x_{\rm c} 
- \frac{32 \pi}{5} x_{\rm c}^{3/2} \right],
\ea
which then renders Eq.~\eqref{Fdot-Eq} into
\ba
\dot{F} &=&  \frac{48}{5 \pi c^{5}} \left(G_{\rm c} {\cal{M}}\right)^{-2} x^{11/2} \; \eta^{11/5}
\nonumber \\
&\times& 
\left\{1 - \left( \frac{743}{336} + \frac{11}{4} \eta \right) x + 4 \pi x^{3/2}  
\right.  
\\ \nonumber
&+& \left.
\frac{65}{768} \dot{G}_{\rm c} {\cal{M}} \; x^{-4} \eta^{-8/5}
\left[1 - \left( \frac{743}{13104} + \frac{11}{156} \eta \right) x \right]
\right\}.
\ea
%

The promotion of $G$ to a time-function with $M$ constant is similar to the promotion of $M$ to a time-function with $G$ constant. 
One can establish this similarity by noting that the evolution of the GW luminosity function and binding energy is governed by the PN
quantity $x$, which is always given in terms of the product $G \; M$. If one were to assume that $M$ varied but not $G$, one would
obtained the same type of evolution equation for the frequency [Eq.~\eqref{Fdot-Eq}], but with a slightly different prefactor.  
One can then qualitatively think of a varying $G$ as a varying $M$, or equivalently a varying chirp mass. 

The differential equation in Eq.~\eqref{Fdot-Eq} can be solved perturbatively in $\dot{G}_{\rm c} {\cal{M}} \ll 1$ up to errors of 
${\cal{O}}(\dot{G}_{\rm c}^{2} {\cal{M}}^{2})$:
\begin{widetext}
\ba
t(F) &=& t_{c}  - \frac{5}{256} G_{\rm c} {\cal{M}} \; \eta^{-8/5} \; x_{\rm c}^{-4}
\left[ 1 + \left(\frac{743}{252} + \frac{11}{3} \eta \right) x_{\rm c} - \frac{32 \pi}{5} x_{\rm c}^{3/2}
\right. 
\nonumber \\
&-& \left. 
\frac{65}{1536} \dot{G}_{\rm c} {\cal{M}} \; \eta^{-8/5} \; x^{-4}_{\rm c} 
- \frac{55}{768} \left(\frac{743}{252} + \frac{11}{3} \eta \right) \dot{G}_{\rm c} {\cal{M}} \; \eta^{-8/5} \; x_{\rm c}^{-3}
+ \frac{5 \pi}{12} \dot{G}_{\rm c} {\cal{M}} \; \eta^{-8/5} \;  x_{\rm c}^{-5/2} 
\right],
\label{tofF}
\ea
\end{widetext}
where $x_{\rm c} = x$ is evaluated at $G(t) = G_{\rm c}$. 
One might also expect corrections to quadrupole emission 
(such as dipolar or monopolar emission) due to a varying Newton's constant. 
Such non-quadrupolar emission could only be sourced by energy carried away by the $G$ field. 
However, if we expect modified $G$-theories to still have well-defined, conserved energies, 
then the effective energy associated with $G$ would scale as $\dot{G}_{\rm c}^2$, which is sub-leading. 
If this behavior is generic, we can ignore such corrections, and we do so henceforth.

Note that the PN equations are ill-behaved at the time of coalescence $t_c$. Though we will
stop the integration of the equations before this time, one might be concerned
about the validity of Taylor expanding Newton's constant about $t_c$ as in
Eq.~\eqref{parametrization}. This may have been an issue if we had taken into account possible back-reaction 
effects of the orbital motion onto the local value of $G$, which we are not, as just discussed.
Hence, to this order, it is valid to picture $G(t)$ as an independent, cosmological field, 
whose functional behavior is not tied to the orbital motion of the binary.

A modification to the frequency directly affects the GW phase evolution and thus the response function.
The Fourier transform of this function in the stationary-phase approximation~\cite{Yunes:2009yz} 
becomes $\tilde{h}(f) = A(f) \exp(i \Psi)$, with
\begin{widetext}
\ba
A(f) &=& - \frac{\sqrt{30 \pi}}{48}  \frac{(G_{\rm c} {\cal{M}}_{z})^{2}}{D_{L}} Q(\iota, \beta) u^{-7/6}  
\left\{1 -\frac{5}{512} \dot{G}_{\rm c} {\cal{M}}_{z} \; u^{-8/3} 
\left[ 1 + \left( \frac{3715}{6048} + \frac{55}{72} \eta \right) \eta^{-2/5} u^{2/3} + 2 \pi \eta^{-3/5} u  \right]
\right\}
\label{newamp}
\\
\Psi &=& 2 \pi f t_{c} - \phi_{c} + \frac{3}{128} u^{-5/3} \left\{
1 + \left( \frac{3715}{756} + \frac{55}{9} \eta \right) \eta^{-2/5} u^{2/3} - 16 \pi \eta^{-3/5} u  
\right.
\nonumber \\
&-& \left. \frac{25}{1536} \dot{G}_{\rm c} {\cal{M}}_{z} u^{-8/3}\left[1 
+ \left( \frac{743}{126} + \frac{22}{3} \eta \right) \eta^{-2/5} u^{2/3}
- \frac{64 \pi}{5} \eta^{-3/5} u\right] \right\},
\label{newphase}
\ea
\end{widetext}
where $u \equiv \pi G_{\rm c} {\cal{M}}_{z} f$ is a dimensionless parameter with $f$ the GW frequency, 
${\cal{M}}_{z} = (1 + z) {\cal{M}}$ is the redshifted chirp mass, and $\phi_{c}$ is a constant phase of coalescence.
$Q(\iota,\beta)$ is some function of the inclination angle $\iota$ and the polarization
angle $\beta$, which depends on the beam-pattern functions~\cite{Yunes:2009yz}.
In Eq.~\eqref{newamp}, we have omitted all the known PN amplitude corrections of GR and we have only explicitly 
written out the modifications to the amplitude induced by a time-varying $G$. 

The frequency dependence of the $\dot{G}_{c}$ modification ($f^{-13/3}$) is unique within the space of 
known alternative theories of gravity~\cite{Yunes:2009ke}. Notice that the controlling factor of the expansion 
is changed by the $\dot{G}_{c}$ terms from $u^{-5/3}$ to $u^{-13/3}$, which implies that to obtain corrections proportional to positive 
powers of the reduced frequency $u$, one would need to include corrections induced by $4$PN terms. 
Comparing Eq.~\eqref{newphase} to the analogous Brans-Dicke modification, we first notice that the $\dot{G}_{\rm c}$ term 
couples to $M \; \eta^{-2}$ to leading order instead of to a positive power
of the symmetric mass ratio $\eta$, as in Brans-Dicke theory. In the latter,
dipolar radiation activates when there are differences in the gravitational and inertial centers of mass (WEP violation), which requires
objects with different self-gravitational binding energies (sensitivities) and mass ratios. On the other hand, the $\dot{G}_{\rm c}$ modification
can be present in the absence of matter (ie.~it modifies the point-particle 
representation of a BH spacetime).  In spite of these differences, the leading-order $\dot{G}_{\rm c}$ modification can be recovered by the simplest 
form of the recently proposed parameterized post-Einsteinian framework~\cite{Yunes:2009ke}, with the leading-order 
post-Einsteinian parameters
\ba
\alpha &=& - \frac{5}{512} \frac{\dot{G}_{\rm c}}{G_{\rm c}} (G_{\rm c} {\cal{M}}_{z}),
\qquad
a = -\frac{8}{3},
\nonumber \\
\beta &=& - \frac{25}{65536} \frac{\dot{G}_{\rm c}}{G_{\rm c}} (G_{\rm c} {\cal{M}}_{z}),
\qquad
b = -\frac{13}{3}.
\ea
%

\section{Placing a Bound on $\dot{G}$}
\label{sec_cons}

With the insight gained from the analytic analysis of the previous section, we shall now concentrate on 
producing estimates of the accuracy to which $\dot{G}_{c}/G_{c}$ could be measured with a LISA GW observation.
In Sec.~\ref{acc-estimates} we briefly describe the basics of a dephasing accuracy estimate and of a Fisher estimate.  In Sec.~\ref{PN-estimates}, we then proceed to calculate estimates using the PN expansion found in Sec.~\ref{sec_mod}.
The resultant expressions, computed using only the leading order PN terms, give insight
into the scaling behavior of the estimates with system parameters. 
More accurate estimates can be obtained if one includes higher order PN terms in the phasing.
In Sec.~\ref{EOB-graphical} we compute the bounds on a $\dot{G}_{c}/G_{c}$ measurement 
using numerically generated waveforms that employ a certain resummation of the PN equations of motion and include 
higher order PN terms. We discuss all the different estimates in Sec.~\ref{comp-estimates}.

\subsection{Accuracy Estimates}
\label{acc-estimates}

One criterion that can be used to estimate the accuracy to which $\dot{G}_{\rm c}$ can be measured is
a {\emph{dephasing measure}}. This measure requires that one compute the GW phase difference between
a waveform with $\dot{G}_{\rm c} = \dot{G}_{0}$ and $\dot{G}_{\rm c} =0$. When this phase difference is 
equal to the inverse of the SNR $1/\rho$, then one estimates that a $\dot{G}_{c}$ correction of magnitude 
$\dot{G}_{0}$ or larger could be {\emph{detected}}. Such a measure follows from the analysis of~\cite{Lindblom:2008cm,Lindblom:2009ux}.

A more popular measure for determining the accuracy to which parameters could be estimated given a GW 
observation is via a Fisher analysis. Consider then the Fourier transform of the response function 
$\tilde{h}(f,\lambda^{a})$ as a frequency-series that depends on certain parameters $\lambda^{a}$ 
that characterize the system. In our case, these parameters define a $6$-dimensional
space, as we have neglected spin, that is spanned by $\lambda^{a} = (A_{0},t_{c},\phi_{c},{\cal{M}},\eta,\dot{G}_{c})$, 
where $A_{0}$ is the first term of Eq.~\eqref{newamp} without the $f^{-7/6}$ factor.
In reality, there are additional (so-called extrinsic) parameters that also enter the response function and characterize 
the motion of the detector around the Sun through time-dependent beam-pattern function. One of the approximations we 
shall make here is to average over all angles so that the detector's proper motion can be ignored.

In the Fisher scheme, the accuracy to which a template parameter $\lambda^{n}$ can be measured is given 
by $\Delta \lambda^{n} = \sqrt{\Sigma^{nn}}$, where $\Sigma^{nn}$ is the $(n,n)$ element of the covariance matrix
(no summation over $n$ implied), defined as the inverse of the Fisher matrix
\be
\Gamma_{ab} = 4 \; \Re \int_{f_{\rm lo}}^{f_{\rm hi}} \frac{df}{S_{n}(f)} \frac{\partial \tilde{h}}{\partial \lambda^{a}} 
\frac{\partial \tilde{h}^{*}}{\partial \lambda^{b}} ,
\ee
with the asterisk denoting complex conjugation. 
The quantity $S_{n}(f)$  is the spectral noise density, which describes the noise in the observation. We 
shall here employ the $S_{n}$ of~\cite{Berti:2004bd}, which includes white-dwarf confusion noise. 
The upper limit of integration $f_{\rm hi}$ is chosen to be the minimum of the 
innermost stable circular orbit (ISCO) frequency $(6^{3/2} \pi M_{z})^{-1}$ and $1 \, {\rm{Hz}}$, while for the lower limit of 
integration $f_{\rm lo}$ we choose the maximum of the frequency $1$ year prior to reaching the ISCO and 
$10^{-5} \; {\rm{Hz}}$ (see eg.~\cite{Berti:2004bd} for further details). 

Another approximation we make here is to ignore correlations between $\dot{G}_{c}$ and other parameters. 
With this approximation at hand, the Fisher matrix becomes diagonal and its inverse tells us how well
parameter $\lambda^{n}$ can be measured: $\Delta \lambda^{n} \sim \left(\Gamma_{nn} \right)^{-1/2}$. 
In our case, if no $\dot{G}_{c}$ modification is observed, we should be able to place a constraint of approximately $\Delta \dot{G}_{c} \equiv \dot{G}_{c}/G_{c} \lesssim \left(\Gamma_{\dot{G}_{c} \dot{G}_{c}} \right)^{-1/2}$. 
The estimate obtained from $\sqrt{\Sigma^{nn}}$, with the considerations described above, is what we refer to as a {\emph{simplified Fisher analysis}} in this paper. 

Correlations generically tend to weaken the accuracy to which parameters can be measured. In our case, the inclusion of the mass ratio 
seems to deteriorate the constraint by less than one order of magnitude. Other physical effects, such as spin and eccentricity could also 
deteriorate the constraint, but it is unclear whether this is the case. This is because such effects appear at high PN order, multiplied by a 
high power of the GW frequency and become important in the late inspiral, while the $\dot{G}_{\rm c}$ corrections modify the leading-order 
Newtonian term, multiplied by a different power of the GW frequency, and dominate during the very early inspiral. 
Only a full Fisher analysis will reveal how much degeneracies deteriorate the constraint, but we relegate such considerations to future work.

\subsection{Stationary Phase, $1.5$ PN Estimates}
\label{PN-estimates}

Let us now apply these measures to the frequency-domain PN waveform found in Sec.~\ref{sec_mod}. 
We begin with the dephasing measure and make use of the fact that the GW frequency to leading PN order 
and zeroth-order in $\dot{G}_{c}$ is given by
\be
f = \frac{5^{3/8}}{8 \pi} \; {\cal{M}}_{z}^{-5/8} \; T^{-3/8} + {\cal{O}}(\dot{G}_{c}),
\ee
where $T$ is the integration time. Inserting this into Eq.~\eqref{newphase}, retaining only the leading-order
$\dot{G}_{c}$ term and requiring that $\Psi(\dot{G}_{c}) - \Psi(\dot{G}_{c}=0) = 1/\rho$ leads to the bound
\be
\Delta \dot{G}_{c} \lesssim \frac{8}{5^{3/8}} \; \frac{1}{\rho} \; \eta^{3/8} \; M_{z}^{5/8} \; T^{-13/8},
\ee
where $M_{z} = M (1 + z)$ is the redshifted total mass.
In turn, this bound leads to the estimates
\ba
\left(\Delta \dot{G}_{c}\right)_{\rm EMRI} &\lesssim& \frac{3 \times 10^{-8}}{\rm yr} \; \left(\frac{100}{\rho}\right) 
\; \left(\frac{\eta}{10^{-5}}\right)^{3/8} 
\nonumber \\
&\times& \left(\frac{M_{z}}{10^{6} M_{\odot}}\right)^{5/8} \left(\frac{T}{1 \rm yr}\right)^{-13/8},
\nonumber \\
\left(\Delta \dot{G}_{c}\right)_{\rm SMBH} &\lesssim& \frac{1.5 \times 10^{-7}}{\rm yr} \; \left(\frac{1000}{\rho}\right) 
\; \left(\frac{\eta}{1/4}\right)^{3/8} 
\nonumber \\
&\times& \left(\frac{M_{z}}{10^{6} M_{\odot}}\right)^{5/8} \left(\frac{T}{1 \rm yr}\right)^{-13/8}.
\label{T-scaling}
\ea
Clearly, the longer we observe the signal and the higher its SNR, the stronger the bound we can place, eg.~a 
$3$-year observation or an order of magnitude decrease in the total mass gains us roughly an order of magnitude 
on the constraint for each. 

Let us now perform a simplified Fisher analysis with the waveform of Sec.~\eqref{sec_mod}. 
To obtain less complicated expressions where the scaling with system parameters is apparent,
we employ the so-called {\emph{restricted}} PN approximation, where one neglects all amplitude corrections to the 
waveform. We find that the accuracy to which $\dot{G}_{c}$ can be measured is approximately given by
\be
\Delta \dot{G}_{c} \lesssim \frac{65536}{25} \frac{1}{\rho} \frac{\eta^{2}}{M_{z}}  \left(\pi M_{z} f_{0}\right)^{13/3} 
\frac{1}{\sqrt{J_{33}}}
\label{scaling}
\ee
where again we have retained only the leading-order $\dot{G}_{c}$ correction to the GW phase in Eq.~\eqref{newphase}.
In Eq.~\eqref{scaling}, we have defined the reduced moment 
$J_{p} = I_{p}/I_{7}$  and the moment of the distribution~\cite{Poisson:1995ef}
\be
I_{p} \equiv \int_{x_{\rm lo}}^{x_{\rm hi}} dx \frac{x^{-p/3}}{S_{n}(x f_{0})},
\ee
with the dimensionless variable $x \equiv f/f_{0}$ and the arbitrary normalization constant $f_{0} = 10^{-4} \; \rm{Hz}$.
Equation~\eqref{scaling} can be used to derive the following scaling rules:
\ba
\left(\Delta \dot{G}_{c}\right)_{\rm EMRIs} &\lesssim& \frac{4 \times 10^{-8}}{\rm yr} \left(\frac{100}{\rho}\right) 
\left(\frac{\eta}{10^{-5}}\right)^{2}
\nonumber \\
&\times&
\left(\frac{M_{z}}{10^{6} M_{\odot}} \right)^{10/3},
\nonumber \\
\left(\Delta \dot{G}_{c}\right)_{\rm SMBH} &\lesssim& \frac{2 \times 10^{-5}}{\rm yr} \left(\frac{1000}{\rho}\right) 
\left(\frac{\eta}{1/4}\right)^{2}
\nonumber \\
&\times&
\left(\frac{M_{z}}{10^{6} M_{\odot}} \right)^{10/3}.
\label{lo-O-bound}
\ea
The scaling of this bound with total mass and mass ratio is different to that of Eq.~\eqref{T-scaling}, 
because we have dropped in Eq.~\eqref{lo-O-bound} the $J_{33}$ dependence, which itself depends
on observation time, total mass and mass ratio. 

It is somewhat surprising that the Fisher and dephasing estimates are 
quite constistent for EMRIs, though much less so for SMBH mergers.
One reason for this is that we have 
neglected sub-leading PN corrections in the frequency-domain 
waveform of Eq.~\eqref{newphase}, which one would expect to have
a much larger effect for equal mass systems.
To check this, we can compute
the sub-leading PN corrections to $\Gamma_{\dot{G}_{c}\dot{G}_{c}}$, which leads to the bound
\ba
\Delta \dot{G}_{c} &\lesssim& \frac{65536}{25} \frac{1}{\rho} \frac{\eta^{2}}{M_{z}}  \left(\pi M_{z} f_{0}\right)^{13/3} 
\frac{1}{\sqrt{J_{33}}}
\left[ 1 
\right. 
\nonumber \\
&-& \left.
\left(\frac{743}{126}- \frac{22}{3} \eta \right) \left( f_{0} M \pi \right)^{2/3} \frac{J_{31}}{J_{33}} 
\right. 
\nonumber \\
&+& \left.
 \frac{64 \pi}{5} \left( \pi M f_{0}\right) \frac{J_{30}}{J_{33}} \right].
 \label{hi-O-bound}
\ea
For a system with the parameters $M_{z}=10^6 M_{\odot}$, $\eta=1/4$ and $\rho=1000$, 
Eq.~\eqref{hi-O-bound} gives $\Delta \dot{G}_{c} \lesssim 8 \times 10^{-7} \; {\rm{yr}}^{-1}$, which is
more than an order of magnitude stronger than that obtained in Eq.~\eqref{lo-O-bound}. This result
gives us a sense of the importance of the inclusion of higher-order corrections to the GW phase and it indicates
that the SMBH bound in Eq.~\eqref{lo-O-bound} may be unreliable (note that due
to the $\eta$ dependence in (\ref{hi-O-bound}) the EMRI estimate is largely 
unaffected). 
With this in mind, the dephasing and 
simplified Fisher measures suggest that roughly the same order of magnitude constraint on $\dot{G}_{c}$ from
either EMRIs or SMBHs could be obtained by LISA.

\subsection{Effective-One-Body Estimates}
\label{EOB-graphical}

In the previous subsection we have seen that the inclusion of higher-order corrections can influence the
estimated accuracy to which $\dot{G}_{c}$ can be measured. For this reason, in this subsection 
we calculate a more accurate GW phase, obtained via numerical methods through the so-called effective-one-body (EOB) formalism~\cite{Buonanno99,Buonanno00,Damour00,Nagar:2006xv,Damour:2007cb,Damour:2007vq,Damour:2007yf,Damour2007,Buonanno:2007pf,Damour:2008gu,Damour:2008te,Boyle:2008ge,Damour:2009kr,Buonanno:2009qa}. 
Essentially, one numerically evolves the bodies' trajectories by solving the Hamilton-Jacobi equations for a certain resummed, 
high-order PN Hamiltonian. The waveform is then constructed via another resummed PN expression that depends on the 
evolved trajectories. We shall not here review the EOB framework, but instead we follow the prescription described in detail in~\cite{2009GWN.....2....3Y}. In particular, we employ the uncalibrated $5.5$PN order EOB model, including $\eta$-deformations of~\cite{Yunes:2009ef,2009GWN.....2....3Y}, 
although we find our results are rather insensitive to the specific EOB model chosen.

The $\dot{G}_{\rm c}$ modification is straightforward to implement in the EOB scheme.
The conservative dynamics (the Hamiltonian) depend on {\emph{reduced}} coordinates and conjugate momenta (denoted
by overhead hats), which are normalized to the total mass and mass ratio $\mu = \eta M$ respectively.  
The $\dot{G}_{\rm c}$ corrections to the Hamiltonian can then be modeled by rescaling certain reduced quantities by $G(t)$
(see eg.~Sec.~III of~\cite{Damour:2009sm})\footnote{Since we study quasi-circular orbits, 
we can neglect the $\theta$ evolution. The Hamilton-Jacobi equations do not acquire overall multiplicative $\dot{G}_{\rm c}$ 
corrections, as this can be cast completely in terms of dimensionless quantities. The only $\dot{G}_{\rm c}$ corrections 
arise via the rescaling of the coordinates and momenta inside the effective Hamiltonian.}:
$t \to \hat{t} \; M G(t)$,
$r \to \hat{r} \; M G(t)$,
$p_{\phi} \to \hat{p}_{\phi} \; M \mu G(t)$.
The dissipative dynamics (the radiation-reaction force) can be modeled by a certain Pade resummation
of the energy flux, which itself is a function of the orbital velocity.
The $\dot{G}_{\rm c}$ corrections to the radiation-reaction force can then be modeled by reinstating the factors of $G$ 
and employing Kepler's third law, which states $v = [G(t) M \omega]^{1/3}$, 
{\emph{ie.}}~the $\phi$-component of this force is rescaled via 
${\cal{F}}_{\phi}(v) \to {\cal{F}}_{\phi}[G(t)^{1/2} v]$
(see eg.~Sec.~IIIb in~\cite{Buonanno00}). In this way, the Hamiltonian and the dissipative force become
explicit functions of time, leading to modified Hamilton-Jacobi equations with new terms proportional
to $\dot{G}_{\rm c}$, whose numerical solution allows us to compute $\dot{G}_{c}$-modified EOB waveforms.

\begin{figure*}[ht]
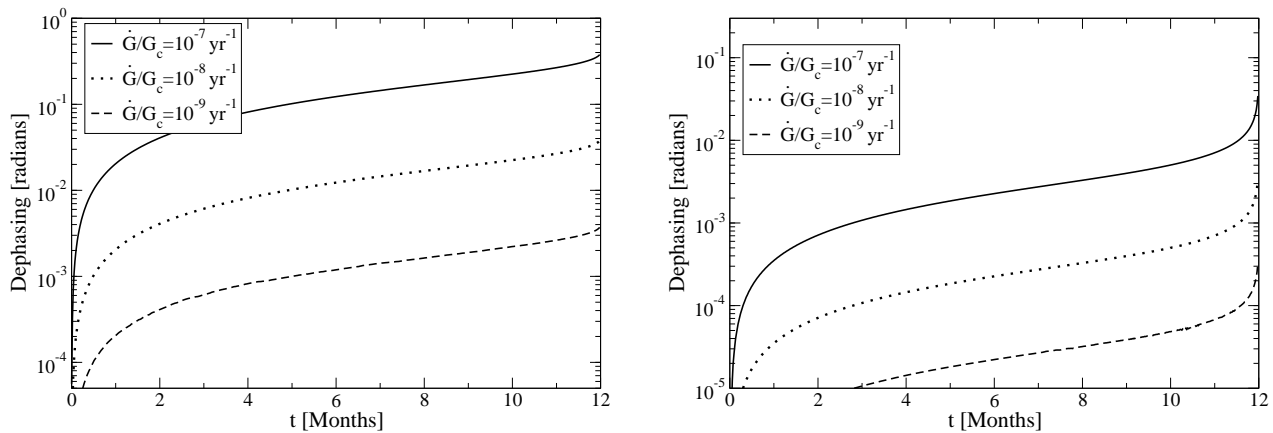

\begin{center}
\begin{tabular}{cc}
 \epsfig{file=phase-diff-EMRI.eps,width=8cm,angle=0,clip=true} & \hspace{0.5cm}
  \epsfig{file=phase-diff-SMBH.eps,width=8cm,angle=0,clip=true} 
\end{tabular}
\end{center}
 \caption{\label{fig.1} Difference in the GW phase between a $\dot{G}_{\rm c} =0$ evolution
and a $\dot{G}_{\rm c}\neq0$  evolution, as a function of time in units of months. Left panel:
fiducial extreme-mass ratio inspiral with $\eta = 10^{-5}$ and redshifted mass $M_{z} := M (1 + z) = 10^{6}$ in the last stages
of coalescence (the orbital radius $r/M \in (6,10)$). Right: fiducial SMBH inspiral $\eta = 1/4$ and 
$M_{z} = 10^{6}$ in the last stages of coalescence (the orbital radius $r/M \in (6,95)$).  }
\end{figure*}

We begin first by applying the dephasing measure to these EOB waveforms.  
Figure~\ref{fig.1} plots the difference in the GW phase
as a function of time in units of months for different values of $\dot{G}_{\rm c}$. 
Clearly, as $\dot{G}_{\rm c}$ is chosen to be larger, the difference in GW phase also increases. With the criteria
described above, these evolutions suggest LISA would be able to detect a 
$\dot{G}_{\rm c} \sim 10^{-9} \; {\rm{yr}}^{-1}$ when observing extreme-mass ratio inspirals (EMRIs) with 
$\rho \sim 250$ or SMBH mergers with $\rho \sim 3300$. As expected from Eq.~\eqref{newphase}, EMRI systems
are preferred over SMBH/SMBH coalescences due to the extra factors of $\eta$ multiplying the $\dot{G}_{\rm c}$
correction (or intuitively, due to the slower inspiral rate, an EMRI spends many more cycles in the
strong-field regime prior to merger, enhancing the secular deviations induced by $\dot{G}_{\rm c}$). 

Let us now apply a numerical version of the simplified Fisher analysis to the modified EOB waveforms. 
Via Parseval's theorem, we can convert a frequency-domain integration into a time-domain integration. 
The Fisher matrix, however, contains the spectral noise density in the denominator, which makes this
conversion non-trivial. We shall here treat this density as an averaged quantity that we evaluate at the 
time $t(f)$. With this in mind, we can rewrite the $(\dot{G}_{c},\dot{G}_{c})$ element of the {\emph{$\rho^{2}$-normalized}} 
Fisher matrix as
\be
\hat\Gamma_{\dot{G}_{c} \dot{G}_{c}} = \frac{4}{\rho^{2}} \int_{0}^{1 \rm{yr}} \frac{dt}{S_{n}[f(t)]} \left| \frac{\partial h}{\partial \dot{G}_{c}} \right|^{2},
\label{time-Gamma}
\ee
and the SNR is simply
\be
\rho^{2} = 4 \int_{0}^{1 \rm{yr}} \frac{dt}{S_{n}[f(t)]} \left| h \right|^{2}.
\ee
The pseudo-averaging of the spectral noise density is similar to what is done in the stationary phase approximation, where the generalized
Fourier integral that defines the Fourier transform is evaluated at the stationary point\footnote{Ref.~\cite{neilpriv} has verified that this time-domain 
implementation of a Fisher analysis produces very similar results to the standard frequency-domain implementation, when the Fourier transform
of the waveform in the latter is treated in a stationary-phase approximation.}.
The accuracy to which $\dot{G}_{c}$ can be measured is then simply 
$\Delta \dot{G}_{c} \lesssim \rho^{-1} (\hat\Gamma_{\dot{G}_{c} \dot{G}_{c}})^{-1/2}$.

We compute the relevant element of the Fisher matrix numerically. The partial derivatives in Eq.~\eqref{time-Gamma}
are approximated via the rule $\partial h/\partial {\dot{G}_{c}} \sim [h(\dot{G}_{0}) - h(0)]/\dot{G}_{0}$ for some very 
small $\dot{G}_{0}$. We have calculated the Fisher estimate for different values of $\dot{G}_{0} \in \left\{10^{-10},10^{-7}\right\} {\rm yr}^{-1}$ 
and found the constraint to be robust to such changes. The waveforms employed are aligned in an initial time window, 
in the low-frequency domain: we have time and phase-shifted them so that they agree to at least one part in $10^{6}$ 
during the first day of evolution. Such a calibration amounts to a matched filtering maximization of the SNR in white noise over 
the time and phase of coalescence~\cite{Buonanno:2009qa} (see~\cite{Yunes:2009ef} for more details on this procedure). 

Using this numerical Fisher prescription and the modified EOB code, we find that for an equal-mass, SMBH inspiral with 
$M_{z} = 10^{6} M_{\odot}$, ${\dot{G}_{c}}$ could be measured to $\Delta {\dot{G}}_{c} \lesssim 3.5 \times 10^{-9} \, {\rm yr}^{-1} (1000/\rho)$. 
Similarly, for an extreme-mass ratio inspiral with the same total mass and mass ratio $\eta = 10^{-5}$,  ${\dot{G}_{c}}$ 
could be measured to $\Delta {\dot{G}}_{c} \lesssim 4.5 \times 10^{-9} \, {\rm yr}^{-1} (100/\rho)$. These estimates are in 
rough agreement with the graphical dephasing measure presented in Sec.~\ref{EOB-graphical}, as well as with the frequency-domain 
simplified Fisher analysis of Sec.~\ref{PN-estimates}, keeping in mind the various approximations employed.

\subsection{Comparing Estimates}\label{comp-estimates}

We have found that the dephasing and simplified Fisher measures are rather consistent among themselves. Differences, however, arise
when considering a Taylor-expanded, stationary-phase approximated, $1.5$PN order waveform versus a modified EOB wave. 
These similarities and differences can be better appreciated in Table~\ref{table}.
\begin{table}[h]
\begin{tabular}{c|c|c|c|c}
\hline\hline
 & PN deph. & PN Fisher & EOB deph. & EOB Fisher \\
\hline
EMRIs & $3 \times 10^{-8}$ & $4 \times 10^{-8}$ & $2.5 \times 10^{-9}$ & $4.5 \times 10^{-9}$ \\
SMBH  & $1.5 \times 10^{-7}$ & $8 \times 10^{-7}$ & $3.3 \times 10^{-9}$ & $3.5 \times 10^{-9}$ \\
\hline\hline
\end{tabular}
\caption{\label{table} Constraints on $\dot{G}_{c}/G$ in units of ${\rm{yr}}^{-1}$, for a 1 year observation, using
the the dephasing (first and third columns) or the Fisher measure (second and fourth columns). 
The first row assumes an EMRI with total redshifted mass $M_{z} = 10^{6}$, symmetric mass
ratio $10^{-5}$ and SNR of $100$, while the second row assumes an equal-mass SMBH inspiral with the 
same total redshifted mass and SNR of $1000$. In the case of the PN Fisher estimate for SMBH
inspirals, we have used the higher-order expressions, as the leading-order term seems unreliable. }
\end{table}
At some level, we expect the EOB waveforms to lead to a better bound, as they include many more $\dot{G}_{c}$ corrections
than the analytic $1.5$PN model. As already found in Sec.~\ref{PN-estimates}, these higher-order PN terms have more
of an impact on comparable mass systems than on EMRIs. 

Overall then, the numerical estimates suggest a {\em typical}
GW inspiral could be used to constrain $\dot{G}_{c}/G_{c}$ to a level of $\sim 10^{-9} \; {\rm{yr}}^{-1}$.
The best bounds come from intermediate mass BH inspirals, eg.~for an inspiral
with redshifted masses $(10^{4}:10^{5}) M_{\odot}$, one can constrain $\dot{G}_{c}/G_{c} \lesssim 10^{-11} \; {\rm{yr}}^{-1}$,
assuming an SNR of $10^{3}$ and an observation time of $3$ years. Such a constraint may even become stronger once spin and eccentricity
are included, as then one can possibly obtain more cycles in the sensitivity band. Realistically, however, 
although the most optimistic systems begin to become comparable to Solar System constraints, 
typical SMBH inspirals at high-redshift will probably not lead to such strong constraints. 
Nevertheless, as explained in the Introduction, a sufficiently large number of observations will allow for the construction of a 
map of the constraints on $\dot{G}$ as a function of redshift and sky position, a task that is impossible 
with Solar System observations. As such, GW tests constitute a new, complementary class of $\dot{G}$ tests, 
allowing for a unique opportunity to build a constraint map during the cosmological epoch of structure formation.

\section{Conclusions}\label{sec_conclusion}

The promotion of Newton's constant to a time-dependent quantity introduces modifications in the GW
frequency and phase of an inspiraling binary. We have analytically computed the $\dot{G}$ modifications to
the response function in the stationary phase approximation to $1.5$PN order.  We have also numerically
solved a $\dot{G}$-modified version of the EOB equations of motion to estimate the magnitude 
of the $\dot{G}$ corrections to higher PN order. We have used both a dephasing and a Fisher measure to obtain 
order of magnitude estimates of the possible constraints that could be placed with LISA given a GW observation. 

An observation of {\emph{one}} merger event could be used to constrain deviations from $G=\rm{const.}$ at the 
spacetime location of merger. A single GW observation of an equal-mass inspiral event with total redshifted 
mass of $10^{5} M_{\odot}$ for $3$ years should be able to constrain $\dot{G}_{c}/G_{c}$ to  $10^{-11} \; {\rm{yr}}^{-1}$. 

Given a sufficient number of GW observations from events at different sky locations and redshifts, one could in
principle construct a {\emph{constraint map}} of $\dot{G}$ as a function of sky angles and redshift. 
Conversely, if $G$ is in truth {\em not} constant, and deviations are measurable, GW observations could allow for a
reconstruction of $G(t,x,y,z)$ along our past lightcone. 
Solar system bounds constrain $\dot{G}$ at redshift zero, while cosmological BBN calculations bound a secular drift in 
$G$ from redshift $0$ to $z \gg 10^3$. With these restrictions, existing bounds are a couple of orders of magnitude stronger
than putative GW observations; however, GWs will allow for entirely new constraints on $\dot{G}$ at intermediate redshifts,
and thus as more events are observed throughout the universe observations will become increasingly sensitive to {\emph{oscillatory}} components in $\dot{G}$ 
(with periods greater than of order a year).

Certainly, the same techniques described here could be used to place constraints on $\dot{G}$ from ground-based
GW observations of solar mass BH mergers with LIGO/VIRGO/GEO. However, the lower expected SNR
of these events, together with the shorter integration time, imply the bounds would be much 
weaker than existing bounds or putative LISA bounds. Furthermore, since the typical, observable 
solar mass merger events will occur at relatively low redshifts, ground-based detectors will also 
not be able to offer a measurement that is complimentary to the already strong solar system constraints.

Future work could concentrate on performing a more detailed data 
analysis of the accuracy to which $\dot{G}_{\rm c}$ can be measured, in particular
taking into account possible degeneracies with other GR binary merger parameters.
We have estimated that such correlations will deteriorate the accuracy to which $\dot{G}$ 
can be constrained by roughly an order of magnitude, when degeneracies with the chirp mass and symmetric mass 
ratio are included, a result in qualitative agreement with~\cite{Berti:2004bd}. The inclusion of additional physical effects, 
such as spin, eccentricity or higher-order PN harmonics, however, might break some of these degeneracies and restore 
some of the strength in the constraint~\cite{Arun:2009pq,Stavridis:2009mb,Yagi:2009zm}. The spin corrections, for example, 
first enter in the phase at $1.5$ PN order via a spin-orbit interaction, and as such are only dominant close to merger, while the
$\dot{G}$ modification also affects the early inspiral. Even if these degeneracies deteriorate the bound dramatically, however,
the GW constraint remains unique in its sensitivity to non-linear behavior in $G$, if a sufficient number of detections are made to 
measure $\dot{G}$ at different locations in the universe.

Another interesting avenue to pursue would be to investigate how the bounds on a varying $G$ could
be strengthened if in addition to the GW signal an electromagnetic (EM) counterpart was observed.
When dealing with pure vacuum spacetimes, such an EM signature is not likely, but recent work on numerical
simulations of mergers embedded in circumbinary disks suggest that EM signals might be 
emitted~\cite{Mosta:2009rr,Bode:2009mt,Palenzuela:2009hx,Anderson:2009fa,Megevand:2009yx,Palenzuela:2009yr}. 
Given a coincident EM/GW detection, one would then be able to determine the position of the source~\cite{Lang:1900bz}
and its redshift~\cite{Hughes:2003ty,Holz:2005df,Cutler:2009qv}, while given a cosmology,
in the form of a distance-redshift relationship, one would then be able to 
independently estimate $G$ at the source. The tandem GW/EM observation would thus simultaneously constraint
$G$ and $\dot{G}$ at the merger event.

One last path to explore is the possible degeneracies between the $\dot{G}_{\rm c}$ effect and the GR time-dependence
of redshift $z$, induced by the expansion of the Universe. Both effects lead to a modification to the GW phase that is proportional 
to the same power of the GW frequency, making these effects exactly degenerate. We performed a preliminary exploration of
such a degeneracy in Appendix~\ref{appendix-cosmo}, where we found that when cosmological information is employed, the $\dot{z}$
effect can be subtracted off to still allow for bounds on 
$\dot{G}$ at the levels discussed in this paper. 

{\bf{Note added after submission}}: Shortly after this paper was submitted, one~\cite{McWilliams:2009ym} appeared 
in preprint considering the possibility of constraining the size of string-theory inspired extra-dimensions, which induce
a time-dependence in the BH mass. As we discussed earlier, such an effect is equivalent to a varying $G$, with the mapping
$\dot{G}_{c}/G \to \dot{M}/M$. Because the extra-dimension effect is strongest for the smallest BHs, we note that EMRIs are the best
systems to explore this effect with GWs. A direct GW detection could place a bound on the size of the extra-dimension of $\ell  \gtrsim 1$ 
micrometers, assuming a constraint on $\dot{M}/M$ on the order of $10^{-9} \; {\rm yr}^{-1}$. Such a constraint can be improved by one 
order of magnitude if the system is observed for more than one year. 

\section{Acknowledgements}
We thank Neil Cornish, Scott Hughes, Sean McWilliams, Leonardo Senatore and Matias Zaldarriaga 
for useful comments. We also want to especially thank Eric Linder for pointing out the cosmological degeneracies 
discussed in Appendix~\ref{appendix-cosmo}.
NY also thanks Jeff Fraser for initially proposing we consider tests of local position invariance. 
 
FP and NY acknowledge support from NSF grant PHY-0745779. 
FP acknowledges the support of the Alfred P. Sloan Foundation.
DS acknowledges support from NASA theory grant NNX087AH30G 
and NSF grant 0707731. 

\appendix
\section{Generic Variable-G Theories and Scalar-Tensor Models}
\label{appendix}

In this appendix, we discuss the relation between the generic variable-G theories we consider in this paper 
and one particular scalar-tensor theory: Brans-Dicke~\cite{1961PhRv..124..925B}.
In Brans-Dicke theory the action is given by
\be
S = \int d^{4}x \sqrt{-g} \left(\frac{\phi}{16 \pi G} R - \frac{\omega_{\BD}}{\phi} \phi_{,a} \phi^{,a} \right) + S_{\rm mat.}, 
\ee
where $g$ is the determinant of the metric, $R$ is the Ricci scalar, $\phi$ is a scalar field, $\omega_{\BD}$ is a 
coupling constant, commas stand for partial differentiation and $S_{\rm mat.}$ stands for additional matter degrees of freedom. 
This action leads to modified field equations, as well as an evolution equation for the scalar field:
\be
\square_{g} \phi = \frac{8 \pi}{3 + 2 \omega_{\BD}} \left( T -  2 \phi \frac{\partial T}{\partial\phi}\right),
\label{eom-phi}
\ee
where $\square_{g}$ is the D'Alembertian operator of curved spacetime and $T$ is the trace of the matter stress-energy tensor. 

Brans-Dicke theory is often interpreted as a variable-$G$ theory by defining a new Brans-Dicke 
gravitational parameter via $G_{\BD} = G \; \phi^{-1}$. 
Thus, if $\phi$ varies in spacetime,
so does the effective coupling parameter $G_{\BD}$. However, note that Newton's constant
$G$ is still exactly that: a constant. What is changing is 
the coupling between other forms of matter and gravity, as mediated by $\phi$. 

In order to illustrate this point,
consider the original weak-field solution found by Brans and Dicke describing the geometry
and scalar field outside a stationary spherically symmetric body of mass $M$\cite{PhysRev.124.925}:
\ba\label{bd_star}
\phi &=& \phi_0 + \frac{2 M}{r} G (3 + 2 \omega_{\BD})^{-1},\nonumber\\
g_{00} &=& -1+\frac{2 M}{r}\frac{G}{\phi_0} [1+(3 + 2 \omega_{\BD})^{-1}],\nonumber\\
g_{ii} &=&  1+\frac{2 M}{r}\frac{G}{\phi_0} [1-(3 + 2 \omega_{\BD})^{-1}],\ \ i\in1,2,3\nonumber\\
g_{ij} &=& 0, \ \ i\ne j
\ea
where $r=\sqrt{x^2+y^2+z^2}$. From the above, the identification of 
an effective $G$ via the action gives
\be
G_{\BD}= \frac{G}{\phi_0 + 2 G M \; r^{-1} \left(3 + 2 \omega_{\BD}\right)^{-1}}.
\ee
Clearly, $G_{\BD}$ varies with distance from the source and it asymptotes
to $G_{\BD} \sim G \; \phi_0^{-1}$ in the limit $M/r \ll 1$. 
However, from the $(0,0)$-component of the metric, the effective Newtonian
potential is $-G_{\rm eff} M/r$, with
\be
G_{\rm eff} = \frac{G}{\phi_0} \left[1+\left(3 + 2 \omega_{\BD}\right)^{-1}\right].
\ee
We notice that this effective potential is 
the same as that of general relativity, but with a {\em constant} rescaling of $G \to G_{\rm eff}$. 
Note further that, in the spatial components of the
metric, a {\em different} constant rescaling of $G$ would bring those
components to a form analagous to Schwarzschild. One of the consequences of this,
as orginally pointed out by Brans and Dicke, is that to this order in the weak-field approximation 
circular orbits are completely
unaffected compared to GR predictions in the solar system, although light deflection is for example
modified.

The example above shows that Brans-Dicke theory is {\em not} fundamentally a variable $G$ theory:
absorbing $\phi$ into an effective variable-$G$ coupling at the level of the action is a formal device and care
must be taken in how this is interpreted.
More significantly, Brans-Dicke theory does not predict any GR deviations in {\em vacuum}. 
In particular, binary BH mergers, as considered in this paper,
are completely unaffected. The reason for this is that Brans-Dicke theory (and all similar scalar-tensor
theories) only alter the coupling between matter stress-energy and gravity; such theories do not change
how the stress-energy of the gravitational field itself back-reacts on the geometry, which
is still controlled by (the unscaled) $G$. 

We can elucidate this further by taking an effective field theory point of view. 
Consider the GR action, linearized in the perturbations $h_{\mu \nu}$ about Minkowski spacetime 
$\eta_{\mu \nu}$, ie.~$g_{\mu \nu} = \eta_{\mu \nu} + h_{\mu \nu}$:
\ba
S &=& \int d^{4}x \left[ \left( h^{\sigma \rho}{}_{,\sigma \rho} - \eta^{\sigma \rho} \square_{\eta} h_{\sigma \rho} \right) 
\right.  
\nonumber \\
&+& \left.
16 \pi G_{\rm matt} {\cal{L}}_{\rm matt} + 16 \pi G_{g} {\cal{L}}_{\GW} 
\right.  
\nonumber \\
&+& \left.
 \alpha^{-1} \left(\partial_{\mu} G_{g}\right) \left(\partial^{\mu} G_{g}\right)\right],
 \label{eff-L}
\ea
where $\square_{\eta}$ is the D'Alembertian operator of flat space, ${\cal{L}}_{\rm matt}$ is some matter Lagrangian density and
${\cal{L}}_{\GW}$ is the second-order, effective Lagrangian density associated with GWs. We have included two coupling constants
here: $G_{\rm{matt}}$ associated with the coupling between gravity and matter; and $G_{g}$ associated with the self-gravitational coupling.
In GR, both of these coupling constants are identical, but in Brans-Dicke $G_{\rm matt}=G/\phi$, whereas
$G_{g}$ is still $G$. Essentially, in Sec.~\ref{sec_mod} of this paper 
we have promoted $G_{g}$ to a function of spacetime, and this cannot be mimicked by Brans-Dicke or similar scalar-tensor theories.

The last term in Eq.~\eqref{eff-L} gives dynamics to the gravitational constant, which leads to the equation of motion
\be
\square_{\eta} G_{g} = \alpha \; {\cal{L}}_{\GW},
\ee
where $\alpha$ is a coupling constant.
Notice that this coupling is different from that which arises in scalar-tensor theories, 
as in the latter the scalar field couples to the trace of the stress-energy tensor, instead of to GW non-linearities. 

\section{Cosmological Degeneracies}
\label{appendix-cosmo}

The modification to the GW response function, derived in Eq.~\eqref{newphase}, is not unique. As mentioned in the Introduction, 
this effect can be also obtained in a theory where $G$ is not varying, but ${\cal{M}}$ is time-dependent. Both of these are 
{\emph{wave generation}} effects, where the evolution equation for the GW frequency is modified due to the new time dependence. 
In addition to this, a similar modification arises if the background cosmology is time-dependent, ie.~in an expanding universe. 
Although such a modification is a {\emph{wave propagation}} one, it leads to the same frequency modification in the frequency-domain 
GW phase to leading order in the stationary-phase approximation.  In this appendix we show how this comes about and consider 
its implications in the constraint of $\dot{G}_{c}$. 

Consider an expanding cosmology, where the metric upon which GWs propagate, is described by an unperturbed 
Friedmann-Robertson-Walker line element: 
\be
ds^{2} = -dt^{2} + a^{2}(t) dx^{i} dx^{j} \delta_{ij},
\ee
where $a(t)$ is the scale factor and $t$ is cosmic time. The scale factor allows us to define the Hubble parameter in the usual
way: $H(t) \equiv \dot{a}/a$, where the overhead dot stands for partial differentiation with respect to cosmic time. Using the standard
definition for the redshift, $z \equiv a_{0}/a -1$, where $a_{0}=a(t_{0})$ corresponds to the present value of the scale factor, we can 
derive the evolution equation
\be
\frac{dz}{dt} = \left(1 + z\right)^{2} H_{0} - \left(1 + z\right) H(z),
\ee
The constant $H_{0}$ is the value of the Hubble parameter $H(z=0)$ today, 
$H_{0} = \left(70.5 \pm 1.3 \right) \; {\rm{km}} \; {\rm{s}}^{-1} \; {\rm{Mpc}}^{-1}$~\cite{Komatsu:2008hk}. 
Notice that this expression differs from that derived by McVittie~\cite{McVittie:1962p13335} 
[$dz/dt_{0} = \left(1 + z\right) H_{0} - H(z)$], because here we differentiate redshift with respect to 
cosmic time at the source, instead of at the observer.

The Hubble parameter depends on the Friedmann equations, and neglecting any $G(t)$ dependence, one can write it as
\be
H(z) = H_{0} \sqrt{\Omega_{m,0} \left(1 + z \right)^{3} + \Omega_{c} + \ldots},
\ee
where $\Omega_{m,0}$ and $\Omega_{c}$ stand for the present-day matter density and the cosmological constant respectively, 
while the $\ldots$ represent other terms that are negligible in the regime $z < 10$.
In a matter-dominated era, the regime of most relevance for LISA, we can approximate the Hubble parameter as 
$H(z) \sim  H_{0} \left(1 + z \right)^{3/2} \Omega_{m,0}^{1/2}$, thus leading to
\be
\frac{dz}{dt} = H_{0} \left[ \left(1 + z\right)^{2} -  \left(1 + z \right)^{5/2} \Omega_{m,0}^{1/2} \right].
\ee
Notice that this effect not only affects gravitational wave observations, but also any electromagnetic one that involves a sufficiently
long, coherent time-integration. 

Let us now connect $\dot{z}$ to the GW response function. As already pointed out, the response function, as measured on Earth,
is sensitive to the {\emph{redshifted}} chirp mass ${\cal{M}}_{z} = (1 + z) {\cal{M}}$. Though since the redshift changes with time, which 
to leading order we can expand about the time of coalescence as
\ba
z(t) &\approx& z_{c} + \left.\frac{dz}{dt}\right|_{z=z_{c}} \left(t - t_{c} \right),
\label{zoft}
\\ \nonumber 
&\approx& z_{c} + H_{0} \left[ \left(1 + z_{c}\right)^{2} -  \left(1 + z_{c} \right)^{5/2} \Omega_{m,0}^{1/2} \right] \left(t - t_{c} \right),
\ea
one sees that ${\cal{M}}_{z}$ becomes time-dependent. 
Such an effect contributes to any cosmological observable
that depends on redshift, as first pointed out by McVittie~\cite{McVittie:1962p13335}. 

The calculation of the GW observable is not yet complete as one must relate the time dependence $(t - t_{c})$
in Eq.~\eqref{zoft} to GW frequency. As pointed out in Sec.~\ref{sec_mod}, the Fourier transform of the GW response
function in the stationary phase approximation is given by Eqs.~\eqref{newamp}-~\eqref{newphase}, except
that now redshift has become a function of time in the inspiral frame. One can rewrite this function of time as a function
of GW frequency, through Eq.~\eqref{tofF} with the substitutions $F \to f/2$ and ${\cal{M}} \to {\cal{M}} (1 + z_{c})$, 
which appropriately redshifts all physical lengths into the observer frame. 

Concentrating on the dominant term in the GW phase, and defining $\Psi_{\rm l.o.}$ to be the leading order term of Eq.~\eqref{newphase} and $\delta \Psi_{\dot{z}}$
to be the new contribution to Eq.~\eqref{newphase}, such that $\Psi = \Psi_{\rm l.o.} (1 + \delta \Psi_{\dot{z}})$, we find
\be
\delta\Psi_{\dot{z}} = \frac{25}{768} \dot{z}_{c} \left(1 + z_{c}\right)^{-1}\; G_{c} {\cal{M}}_{z_{c}} \; u^{-8/3}\,,
\label{deltapsicosmo}
\ee
where now $u \equiv \pi G_{c} {\cal{M}}_{z_{c}}f$ and ${\cal{M}}_{z_{c}} \equiv {\cal{M}} (1 + z_{c})$.
The new relative contribution, $\delta \Psi_{\dot{z}}$, depends on $f^{-8/3}$, or equivalently, $u^{-8/3}$, 
just as the $\dot{G}_{c}$ term in Eq.~\eqref{newphase}, and thus
the two effects are degenerate. 

This raises the question of whether the $\dot{z}$ effect could overwhelm
the $\dot{G}_{c}$ effect, 
rendering the bounds derived in this paper unattainable. 
To answer this question, we will use the results of the PN dephasing/Fisher
analysis of the $\dot{G}_{c}$ effect in Sec.~\ref{PN-estimates}.
Using the measured values of the cosmological parameters obtained in the
five-year WMAP analysis~\cite{Dunkley:2008ie,Komatsu:2008hk} 
($H_{0} = 70.5 \; {\rm{km}} \; {\rm{s}}^{-1} \; {\rm{Mpc}}^{-1}$, $\Omega_{m,0} = 0.26$),
and defining the ratio $r \equiv \delta \Psi_{\dot{z}}/\delta \Psi_{\dot{G}_{c}}$ of the $\delta\Psi_{\dot{z}}$ 
correction induced by the $\dot{z}$ term to the $\delta \Psi_{\dot{G}_{c}}$ correction induced by the 
$\dot{G}_{c}$ term (the first term in the second line of Eq.~\eqref{newphase}),
we find that $r \approx 0.008 \; \; (10^{-8} \; {\rm{yr}}^{-1}/\dot{G}_{c})$ at $z_{c}=1$, growing to
$r \approx 0.02\; \; (10^{-8} \; {\rm{yr}}^{-1}/\dot{G}_{c})$ at $z_{c}=5$ for equal-mass binaries.
Since this ratio is less than one, it implies the cosmological effect is
not significant, and does not need to be accounted for. 

One might wonder
though why we did not use the more sensitive bounds for $\dot{G}_{c}$ 
obtained from the EOB analysis in the above. The reason is that the $\dot{z}$
effect is purely a propagation effect, and would not be enhanced by
inclusion of higher-order PN terms. This is in contrast to the
effect of $\dot{G}_{c}$, which changes the local dynamics of the binary,
and for near-equal mass systems this is strongly reflected in the higher
order PN terms.
Nevertheless, even if we did {\emph{assume}} the EOB $\dot{G}_{c}$ bounds
carried over verbatum to $\dot{z}$, one can still subtract the
$\dot{z}$ effect using cosmological data prior to testing for $\dot{G}_{c}$. 
How much this will reduce the $\dot{z}$ effect depends on how accurately
the cosmology is known. Assuming a relative error of $10\%$
in $H_{0}$ and $\Omega_{m,0}$ (consistant with how accurately
these numbers are known today), the ratio is reduced 
to $r \approx 0.2 \; \; (10^{-10} \;{\rm{yr}}^{-1}/\dot{G}_{c})$ at $z = 1$ and $r \approx 0.5 \; \; (10^{-10} \;{\rm{yr}}^{-1}/\dot{G}_{c})$ at z = 5.
From new and anticipated observations, provided for example
by Planck~\cite{:2006uk}, it is not unreasonable to assume by the
anticipated time of LISA the relative error in the cosmology
would be an order of magnitude less, further reducing $r$ by
an order of magnitude. This is then well below our most optimistic
bounds for measuring $\dot{G}_{c}$.

\bibliographystyle{apsrev}
\bibliography{master}

\end{document}